\documentclass[prc,aps,twocolumn,showpacs,preprintnumbers,amsmath,amssymb]{revtex4}
\usepackage{graphicx}
\usepackage{psfig}
\usepackage{dcolumn}

\begin{document}
\title{Evidence for nonhadronic degrees of freedom in the transverse mass
spectra of kaons from relativistic nucleus-nucleus collisions?}
\author{E.~L.~Bratkovskaya}
 \affiliation{Institut f\"{u}r Theoretische Physik,
   Universit\"{a}t Frankfurt, 60054 Frankfurt, Germany}
\author{M. van Leeuwen}
 \affiliation{NIKHEF, Amsterdam, Netherlands}
\author{S. Soff}
 \affiliation{Institut f\"{u}r Theoretische Physik,
   Universit\"{a}t Frankfurt, 60054 Frankfurt, Germany}
\author{W. Cassing}
 \affiliation{Institut f\"{u}r Theoretische Physik,
   Universit\"{a}t Giessen, 35392 Giessen, Germany}
\author{H.~St\"ocker}
 \affiliation{Institut f\"{u}r Theoretische Physik,
   Universit\"{a}t Frankfurt, 60054 Frankfurt, Germany}

\begin{abstract}
We investigate transverse hadron spectra from relativistic
nucleus-nucleus collisions which reflect important aspects of the
dynamics - such as the generation of pressure  - in the hot and dense
zone formed in the early phase of the reaction. Our analysis is
performed within two independent transport approaches (HSD and UrQMD)
that are based on quark, diquark, string and hadronic degrees of
freedom.  Both transport models show their reliability for elementary
$pp$ as well as light-ion (C+C, Si+Si) reactions.  However, for central
Au+Au (Pb+Pb) collisions at bombarding energies above $\sim$ 5
A$\cdot$GeV the measured $K^{\pm}$ transverse mass spectra have a
larger inverse slope parameter than expected from the calculation. Thus
the pressure generated by hadronic interactions in the transport models
above $\sim$ 5 A$\cdot$GeV is lower than observed in the experimental
data. This finding shows that the additional pressure - as expected
from lattice QCD calculations at finite quark chemical potential and
temperature - is generated by strong partonic interactions in the early
phase of central Au+Au (Pb+Pb) collisions.
\end{abstract}

\maketitle

The phase transition from partonic degrees of freedom (quarks and
gluons) to interacting hadrons is a central topic of modern high-energy
physics. In order to understand the dynamics and relevant scales of
this transition laboratory experiments under controlled conditions are
presently performed with ultra-relativistic nucleus-nucleus collisions.
Hadronic spectra and relative hadron abundancies from these experiments
reflect  important aspects of the dynamics in the hot and dense zone
formed in the early phase of the reaction.
Furthermore, as has been proposed early by Rafelski and M\"uller
\cite{Rafelski1} the strangeness degree of freedom might play an
important role in distinguishing hadronic and partonic dynamics.

Estimates based on the Bjorken formula \cite{bjorken} for the energy
density achieved in central Au+Au collisions suggest that the critical
energy density for the formation of a quark-gluon plasma (QGP) is by
far exceeded during a few fm/c in the initial phase of the collision at
Relativistic Heavy Ion Collider (RHIC) energies \cite{QM01}, but
sufficient energy densities ($\sim$ 0.7-1 GeV/fm$^3$ \cite{Karsch})
might already be achieved at Alternating Gradient Synchrotron (AGS)
energies of $\sim$ 10 $A\cdot$GeV \cite{HORST,exita}. More recently,
lattice QCD calculations at finite temperature and quark chemical
potential $\mu_q$ \cite{Fodor} show a rapid increase of the
thermodynamic pressure $P$ with temperature above the critical
temperature $T_c$ for a phase transition to the QGP. The crucial
question is, however, at what bombarding energies the conditions  for
the phase transition are fulfilled.

Presently,  transverse mass (or momentum) spectra of hadrons are in the
center of interest: the measured transverse
mass $m_T$ spectra of kaons \\
\vspace*{-5mm}
\begin{eqnarray}
\label{slope}
\frac{1}{m_T} \frac{dN}{dm_T} \sim \exp(-\frac{m_T}{T})
\end{eqnarray}
at  AGS and SPS (from 30 to 160 $A\cdot$GeV at the Super Proton
Synchrotron) energies  show  a substantial {\it flattening} or {\it
hardening} of the spectra in central Au+Au collisions relative to $pp$
interactions (cf. \cite{NA49_T,Goren}).  This hardening of the spectra
(or increase of the inverse slope parameter $T$ in (\ref{slope})) is
commonly attributed to strong collective flow which is absent in the
respective $pp$ data. We will demonstrate in this Letter that the
pressure needed to generate such a large collective flow cannot be
produced by the interactions of hadrons in the expansion phase of the
hadronic fireball.

Two independent transport models that employ hadronic and string
degrees of freedom, i.e. UrQMD \cite{UrQMD1,UrQMD2} and HSD
\cite{Geiss,Cass99}, are used for our study. They take into account the
formation and multiple rescattering of hadrons and thus dynamically
describe the generation of pressure in the hadronic expansion phase.
The UrQMD transport approach \cite{UrQMD1,UrQMD2} includes all baryonic
resonances up to  masses of 2 GeV as well as mesonic resonances up to
1.9 GeV as tabulated by the Particle Data Group \cite{PDG}. For
hadronic continuum excitations a string model is used with hadron
formation times in the order of 1-2~fm/c depending on the momentum and
energy of the created hadron.  The transport approach is matched to
reproduce the nucleon-nucleon, meson-nucleon and meson-meson cross
section data in a wide kinematic range \cite{UrQMD1,UrQMD2}.
In the HSD approach nucleons, $\Delta$'s, N$^*$(1440), N$^*$(1535),
$\Lambda$, $\Sigma$ and $\Sigma^*$ hyperons, $\Xi$'s, $\Xi^*$'s and
$\Omega$'s  as well as their antiparticles are included on the baryonic
side whereas the $0^-$ and $1^-$ octet states are included in the
mesonic sector. High energy inelastic hadron-hadron collisions in HSD are
described by the FRITIOF string model \cite{LUND} whereas low energy
hadron-hadron collisions are modeled based on experimental cross
sections. We point out importantly, that no parton-parton scattering
processes are included in the studies below contrary to the multi-phase
transport model (AMPT) \cite{Ko_AMPT}, which is currently employed from
upper SPS to RHIC energies.

Whereas the underlying concepts of the two transport theoretical
models are very similar, the actual realizations differ
considerably. Indeed, a systematic analysis of results from both models
and experimental data for central nucleus-nucleus collisions from 2 to
160 $A\cdot$GeV has shown that the 'longitudinal' rapidity distributions of
protons, pions, kaons, antikaons and hyperons are quite similar in both
models and in  reasonable agreement with available data.  The exception
are the pion rapidity spectra at the highest AGS energy and lower SPS
energies, which are overestimated by both models \cite{Weber02}.  For a
comparison of HSD calculations with experimental data at RHIC energies
we refer the reader to Ref. \cite{Brat03}.

We focus on transverse mass spectra of pions
and kaons/antikaons from central Au+Au (Pb+Pb) collisions from
2 $A\cdot$GeV to 21.3 $A\cdot$TeV  and compare to recent data.
\begin{figure}[t]
\centerline{\psfig{file=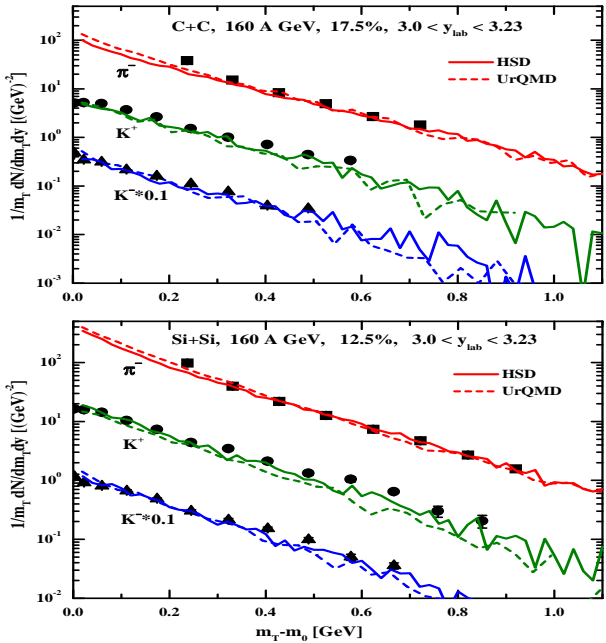,width=8cm,height=8.5cm}}
\vspace*{-3mm}
  \caption{Comparison of transverse mass spectra from HSD (solid lines)
  and UrQMD (dashed lines) for $\pi^-, K^+$ and $K^- (\times 0.1)$ from
  17.5\% central C+C (upper part) and 12.5\% central Si+Si (lower part)
  reactions at 160 $A\cdot$GeV close to midrapidity with data
  from Ref. \protect\cite{NA49_CCSi} (full symbols). }
  \label{bild2}
\end{figure}
We start with results on $m_T$ spectra for central C+C and Si+Si
collisions at 160 $A\cdot$GeV.  In Fig. \ref{bild2} the experimental $m_T$
spectra for $\pi^-$, $K^+$ and $K^-$ for C+C reactions (17.5\%
centrality; upper part) and Si+Si (12.5\% centrality; lower part)
\cite{NA49_CCSi} are compared to HSD (solid lines) and UrQMD
calculations (dashed lines).  The $m_T$ spectra are calculated in a
rapidity interval (3.0 $\leq y_{\rm lab} \leq $  3.23) close to
midrapidity. The agreement between the transport calculations and the
data is quite satisfactory; no obvious traces of 'new' physics are
visible.
\begin{figure}[t]
\centerline{\psfig{file=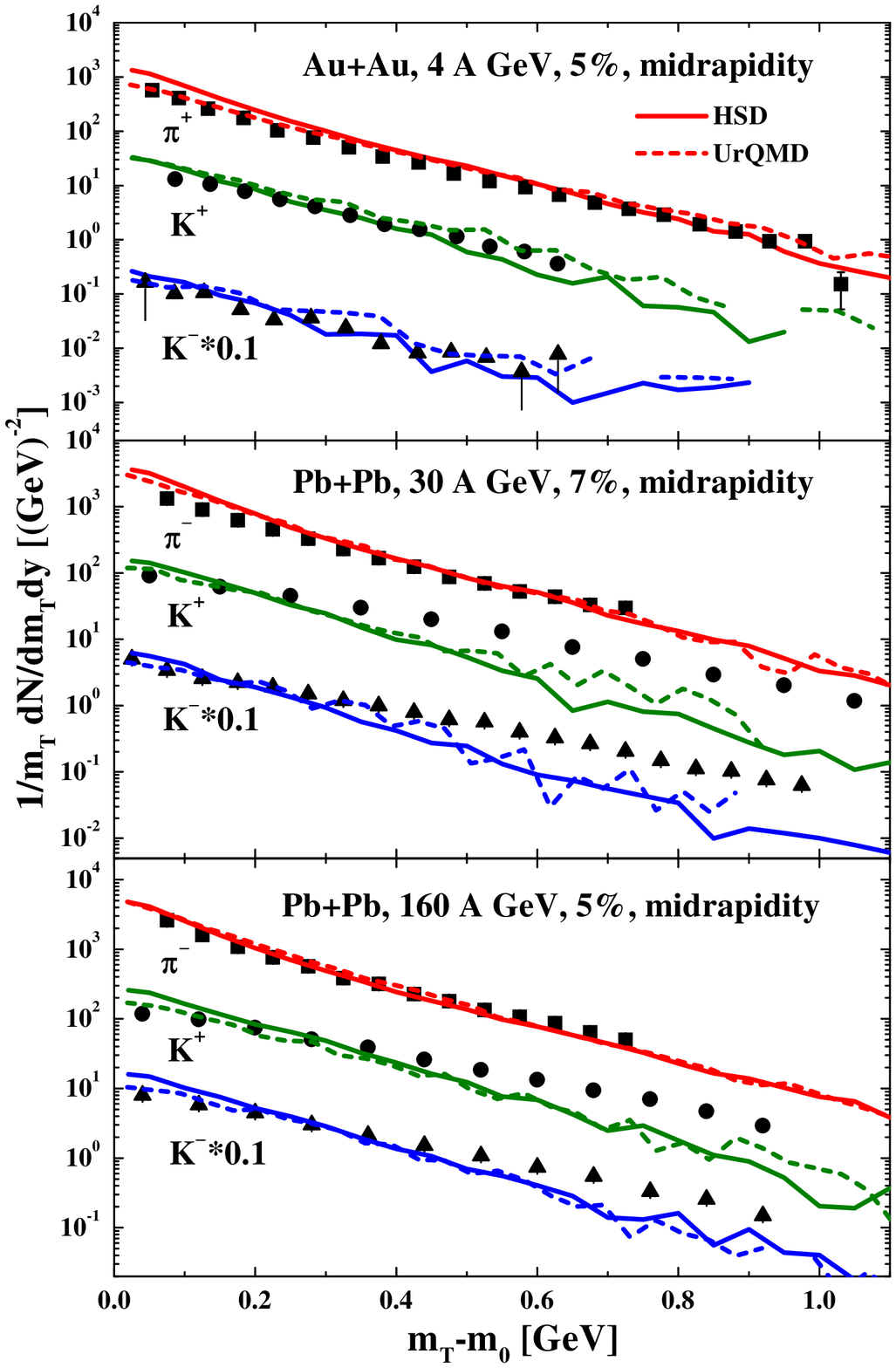,width=8cm,height=9.6cm}}
\vspace*{-3mm}
  \caption{Comparison of transverse mass spectra from HSD (solid lines)
  and UrQMD (dashed lines) for $\pi^\pm, K^+$ and $K^- (\times 0.1)$ from
  central Au+Au (Pb+Pb) reactions at 4, 30, and 160 $A\cdot$GeV at
  midrapidity with the data from Refs. \protect\cite{NA49_T,E866E917}
  (full symbols).}
  \label{bild3}
\end{figure}
The situation, however, changes for central Au+Au (or Pb+Pb) collisions
as demonstrated in Fig. \ref{bild3}, where we display the calculated
spectra from UrQMD (dashed lines) and HSD (solid lines) for 5\%
(or 7 \%) central Au+Au (or Pb+Pb) collisions at 4, 30 and 160
$A\cdot$GeV in comparison to midrapidity data from Refs.
\cite{NA49_T,E866E917}
\footnote{Note that all data from the NA49 Collaboration at 30 A$\cdot$GeV
have to be considered as 'preliminary'.}.
Whereas at the lowest energy of 4 $A\cdot$GeV the agreement between the
transport approaches and the data is still acceptable, severe
deviations are visible in the $K^\pm$ spectra at SPS energies of 30 and
160 $A\cdot$GeV.  Note that the $\pi^{\pm}$ spectra are reasonably
described at all energies while the inverse slope $T$ of the $K^\pm$
transverse mass spectra is underestimated severely by about the same
amount in both transport approaches (within statistics).  The increase
of the inverse $K^\pm$ slopes in heavy-ion collisions with respect to
$pp$ collisions, which is generated by rescatterings of produced
hadrons in the transport models, is only small because the elastic
meson-baryon scattering is strongly forward peaked and therefore gives
little additional transverse momentum at midrapidity.

The question remains whether the discrepancies shown in Fig.
\ref{bild3} might be due to conventional hadronic medium effects.  In
fact, the $m_T$ slopes of kaons and antikaons at SIS energies (1.5 to
2 $A\cdot$GeV) were found to differ significantly \cite{KaoS}. As argued
in  \cite{Cass99} the different slopes could be traced back to
repulsive potentials of kaons with nucleons, which lead to a hardening
of the $K^+$ spectra,  and attractive antikaon-nucleon potentials,
which lead to a softening of the $K^-$ spectra. However,
the effect of such potentials was calculated within HSD and was found
to be of minor importance at AGS and SPS energies \cite{Cass99} since
the meson densities are comparable to or even larger than the baryon
densities at AGS energies and above.

Additional self energy contributions stem from $K^\pm$ interactions
with mesons; however, $s$-wave
kaon-pion interactions are weak due to chiral symmetry arguments and
$p$-wave interactions such as $\pi+K \leftrightarrow K^*$ transitions
are  suppressed substantially by the approximately 'thermal' pion
spectrum. A recent study on the kaon potentials in hot pion matter
gives kaon mass shifts of about $-52$ MeV and vector potentials of
$\sim +49$ MeV \cite{Fuchs} for a pion gas at temperature $T$= 170 MeV.
We have employed even slightly larger $K^\pm$ potentials in dynamical
HSD calculations and achieved a hardening of the $K^\pm$ spectra of
less than 10\%.

Furthermore, we have pursued the idea of Refs.  \cite{Sorge,Bleich}
that the $K^\pm$ spectra could be hardened by string-string
interactions, which increase the effective string tension $\sigma$ and
thus the probability to produce mesons at high $m_T$
\cite{Bleich,Ko_AMPT}.  In order to estimate the largest possible
effect of string-string interactions we have assumed that for two
overlapping strings the string tension $\sigma$ is increased by a
factor of two, for three overlapping strings by a factor of three etc.
Here the overlap of strings is defined geometrically assuming a
transverse string radius $R_s$, which according to the studies in Ref.
\cite{Geiss99} should be $R_s \leq$ 0.25 fm.  Based on these
assumptions (and $R_s$=0.25 fm), we find only a small increase of the
inverse slope parameters at AGS energies, where the string densities
are low.  At 160 $A\cdot$GeV, the model gives a significant hardening of
the spectra by about 15\%, which, however, is still much less than
in the data in Fig.  \ref{bild3}.
\begin{figure}[th]
\centerline{\psfig{file=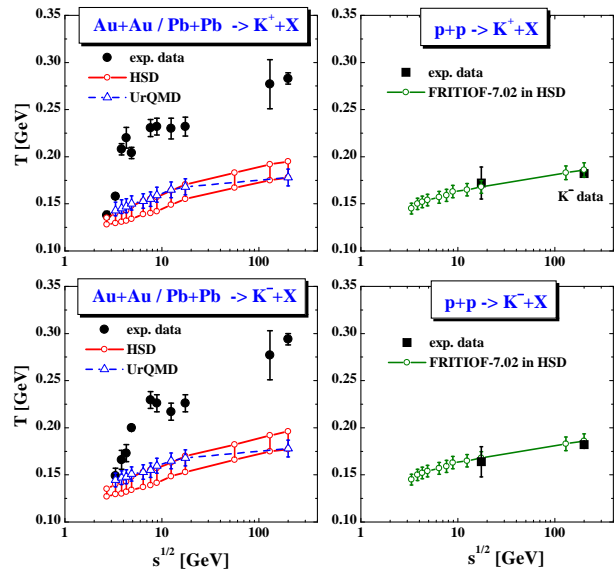,width=8.cm}}
\vspace*{-2mm}
  \caption{Comparison of the inverse slope parameters for $K^+$ and
  $K^-$ mesons from central Au+Au (Pb+Pb) collisions (l.h.s.) and $pp$
  reactions (r.h.s.) as a function of the invariant energy $\sqrt{s}$
  from HSD (upper and lower solid lines) and UrQMD (open triangles)
  with data from Refs. \protect\cite{NA49_T,NA49_CCSi}.
  The upper and lower solid lines result from different limits of
  the HSD calculations as discussed in the text.}
 \label{bild4}
\end{figure}

Our findings are summarized in Fig. \ref{bild4} where the dependence of
the inverse slope parameter $T$ (see Eq.~(\ref{slope})) on $\sqrt{s}$
is shown and compared to experimental data \cite{NA49_T,NA49_CCSi} for
central Au+Au (Pb+Pb) collisions (l.h.s.) and $pp$ reactions (r.h.s.).
The upper and lower solid lines (with open circles) on the l.h.s. in
Fig. \ref{bild4} correspond to results from HSD calculations, where the
upper and lower limits are due to fitting the slope $T$ itself, an
uncertainty in the repulsive $K^\pm$-pion potential or the possible
effect of string overlaps.  The open triangles represent slope
parameters from default UrQMD calculations (version 1.3) which are well
within the limits obtained from the different HSD calculations. It's
worth to point out that in the pure cascade mode the kaon slope $T$ is
slightly larger in the UrQMD calculations than from HSD at AGS and SPS
energies, which is related to the isotropic decay of higher mass ($<$ 2
GeV) resonances in UrQMD  as well as a different string fragmentation
function. The slope parameters from $pp$ collisions (r.h.s. in Fig.
\ref{bild4}) are seen to increase smoothly with energy both in the
experiment (full squares) and in the HSD calculations (full lines with
open circles). The error bars related to our calculations are due to an
uncertainty in extracting the slope parameter when considering
different $m_T$ intervals.

We mention that the RQMD model \cite{Sorge} gives higher inverse slope
parameters for kaons at AGS and SPS energies than HSD and UrQMD, which
essentially can be traced back to the implementation of resonances with
masses above 2 GeV as well as 'color ropes' that decay isotropically in
their rest frame \cite{Hecke}. However, we consider the implementation
of such  'heavy resonances' with unknown properties as a
phenomenological fit to experimental data.

This still leaves us with the question of the origin of the rapid
increase of the $K^\pm$ slopes with invariant energy for central Au+Au
collisions, which is  missed in both transport approaches. We recall
that higher transverse particle momenta either arise from repulsive
self energies -- in mean-field dynamics -- or from collisions, which
reduce longitudinal momenta in favor of transverse momenta.  As shown
in Fig. 3 conventional hadron self-energy effects and hadronic
binary collisions are insufficient to describe the dramatic increase of
the $K^\pm$ slopes as a function of $\sqrt{s}$. We conclude
that this points towards a different mechanism for the
generation of the pressure observed experimentally and propose,
that partonic degrees of freedom should be responsible for this effect
already at $\sim$ 5 $A\cdot$GeV.

Our arguments are based on a comparison of the transverse pressure from
the transport models with recent lattice QCD calculations \cite{Fodor}:
The  transverse pressure ($P_{xx} \approx P_{yy}$) - in the central overlap
volume of Au+Au collisions - within UrQMD is about 0.07 GeV/fm$^3$ at
10.7 A GeV, 0.1 GeV/fm$^3$ at 40 A GeV and 0.15 GeV/fm$^3$ at 160 A GeV
(cf. Ref. \cite{Bravina}).  These numbers should be compared to lattice
QCD calculations.  Adopting the results from Ref. \cite{Fodor}
and using $T_c \approx$ 172 MeV (following Ref. \cite{Rafelski}) we end
up with a pressure (in thermal and chemical equilibrium) of 0.15
GeV/fm$^3$ at vanishing baryon density ($\mu_B$ = 0), 0.17 GeV/fm$^3$
($\mu_B$ = 0.21 GeV), 0.2 GeV/fm$^3$ ($\mu_B$ = 0.33 GeV), 0.22
GeV/fm$^3$ ($\mu_B$ = 0.42 GeV), and 0.26 GeV/fm$^3$  ($\mu_B$ = 0.53
GeV) at a temperature $T=1.1 T_c$ ($\sim$ 190 MeV). These values are
higher than those from the transport calculations, but correspond to
a temperature significantly above $T_c$. The important finding from the
lattice QCD calculations \cite{Fodor} is that the pressure drops by an
order of magnitude when the temperature decreases to $T_c$, which leads
to $\sim$ 0.02-0.03 GeV/fm$^3$ at $T_c$. These numbers now are lower
than the transport results for bombarding energies $\geq$ 10.7
$A\cdot$GeV! Consequently, the  phase transition should be reached even
at a lower bombarding energy  in line with our findings in Fig. 3.

The authors of Ref. \cite{SMES} have proposed to interpret the
approximately constant $K^\pm$ slopes above $\sim 30 A\cdot$GeV as an
indication for a phase transition \cite{SMES}.
However, to support their arguments
more experimental data in the energy range 20 GeV $\leq \sqrt{s}
\leq $ 100 GeV are needed to clarify this issue experimentally.

Summarizing this study, we conclude that according to two independent
string-hadron transport models, HSD and UrQMD, the inverse slope
parameters $T$ for $K^\pm$ mesons are practically independent of system
size from $pp$ up to central Pb+Pb collisions and show a slight
increase with collision energy. The calculated transverse mass spectra
are in reasonable agreement with the experimental results for $pp$
collisions and central collisions of light nuclei (C+C and Si+Si). The
rapid increase of the inverse slope parameters of kaons for collisions
of heavy nuclei (Au+Au) found experimentally in the AGS energy range,
however, is not reproduced by both models (see Fig.~\ref{bild4}).
Since the pion transverse mass spectra are described sufficiently
well at all bombarding energies, the failure has to be attributed to a
lack of pressure especially for the strange $K^\pm$-mesons.
We have argued - based on lattice QCD calculations
for the pressure at finite temperature and baryon chemical potential
$\mu_B$ from Ref. \cite{Fodor} - that this additional pressure must be
generated in the early phase of the collision because the strong
hadronic interactions in the later stages don't produce it.

The interesting finding of our analysis is, that nonhadronic degrees of 
freedom seem to play a substantial role in central Au+Au collisions 
already at AGS energies above $\sim$ 5 $A\cdot$GeV.  Further 
indications for a low energy onset of a phase transition are the missed 
$K^+/\pi^+$ ratios in the AGS and lower SPS regime \cite{Weber02}. 
Additionally, the excitation function of the elliptic flow $v_2$ 
indicates a softening of the equation of state at $\sim$ 5 $A\cdot$GeV 
in Au+Au collisions \cite{Dani,Sahu}, which signals a change of the 
pressure, too.  In spite of the latter observations - all showing a 
significant change at $\sim$ 5 $A\cdot$GeV Au+Au collisions - data for 
weakly interacting (on the hadronic level) $\phi$-mesons  will be 
necessary to clarify further the question raised in the title from the 
experimental side.

The authors like to thank M. Bleicher, C. Greiner, C.M. Ko, Z.W. Lin
and H. Weber for valuable discussions.  E.L.B. was supported by DFG.


\end{document}